\documentstyle[12pt]{article}
\begin{document}


\renewcommand{\a}{\alpha}
\renewcommand{\b}{\beta}
\renewcommand{\d}{\delta}
\newcommand{\pa}{\partial}
\newcommand{\g}{\gamma}
\newcommand{\G}{\Gamma}
\newcommand{\e}{\epsilon}
\newcommand{\z}{\zeta}
\newcommand{\Z}{\Zeta}
\newcommand{\k}{\kappa}
\newcommand{\K}{\Kappa}
\renewcommand{\l}{\lambda}
\renewcommand{\L}{\Lambda}
\newcommand{\m}{\mu}
\newcommand{\n}{\nu}
\newcommand{\x}{\chi}
\newcommand{\X}{\Chi}
\renewcommand{\P}{\Pi}
\newcommand{\r}{\rho}
\newcommand{\s}{\sigma}
\renewcommand{\S}{\Sigma}
\renewcommand{\t}{\tau}
\newcommand{\T}{\Tau}
\newcommand{\y}{\upsilon}
\newcommand{\Y}{\upsilon}
\renewcommand{\o}{\omega}
\renewcommand{\O}{\Omega}
\newcommand{\q}{\theta}
\newcommand{\h}{\eta}
\newcommand{\var}{\varepsilon}

\newcommand\balpha{\mbox{\boldmath $\alpha$}}
\newcommand\bbeta{\mbox{\boldmath $\beta$}}
\newcommand\bgamma{\mbox{\boldmath $\gamma$}}
\newcommand\bomega{\mbox{\boldmath $\omega$}}
\newcommand\blambda{\mbox{\boldmath $\lambda$}}
\newcommand\bmu{\mbox{\boldmath $\mu$}}
\newcommand\bphi{\mbox{\boldmath $\phi$}}
\newcommand\bzeta{\mbox{\boldmath $\z$}}
\newcommand\bsigma{\mbox{\boldmath $\sigma$}}

\newcommand\ualpha{\,\lower 1.2ex\hbox{$\sim$}\mkern-13.5mu {\a}}
\newcommand\ubeta{\,\lower 1.2ex\hbox{$\sim$}\mkern-13.5mu {\b}}
\newcommand\ugamma{\,\lower 1.2ex\hbox{$\sim$}\mkern-13.5mu {\g}}
\newcommand\ug{\,\lower 1.2ex\hbox{$\sim$}\mkern-13.5mu {g}}
\newcommand\ulambda{\,\lower 1.2ex\hbox{$\sim$}\mkern-13.5mu {\l}}
\newcommand\umu{\,\lower 1.2ex\hbox{$\sim$}\mkern-13.5mu {\mu}}
\newcommand\uA{\,\lower 1.2ex\hbox{$\sim$}\mkern-13.5mu {A}}
\newcommand\uphi{\,\lower 1.2ex\hbox{$\sim$}\mkern-13.5mu {\phi}}
\newcommand\uF{\,\lower 1.2ex\hbox{$\sim$}\mkern-13.5mu {F}}
\newcommand\uchi{\,\lower 1.2ex\hbox{$\sim$}\mkern-13.5mu {\chi}}

\newcommand{\Ka}{K\"ahler\ } 
\newcommand{\vhiggs}{{\rm v}}

\newcommand{\ft}[2]{{\textstyle\frac{#1}{#2}}}
\newcommand{\eqn}[1]{(\ref{#1})}
\newcommand{\vsone}{\vspace{1cm}}

\begin{titlepage}
\begin{flushright}
SWAT-98/186 \\
UW/PT 98-1 \\
hep-th/9803148
\end{flushright}
\begin{centering}
\vspace{.2in}
{\large {\bf Three-Dimensional Gauge Theories and ADE Monopoles}}\\
\vspace{.4in}
 David Tong \\
\vspace{.4in}
Department of Physics, University of Wales, Swansea \\
Singleton Park, Swansea, SA2 8PP, UK\\
\vspace{.1in}
and \\
\vspace{.1in}
Department of Physics, University of Washington, Box 351560   \\
Seattle, Washington 98195-1560, USA\\
pydt@fermi.phys.washington.edu\\
\vspace{.4in}
{\bf Abstract} \\
\end{centering}
We study three-dimensional $N=4$ gauge theories with product gauge 
groups constructed from ADE Dynkin diagrams. One-loop corrections 
to the metric on the Coulomb branch are shown to coincide with the metric 
on the moduli space of well-seperated ADE monopoles. We propose that this 
correspondence is exact.


\end{titlepage}

\subsubsection*{\em Introduction}

There exists a correspondence between three-dimensional 
gauge theories and monopole moduli spaces. The Coulomb branch of 
$N=4$ supersymmetric $U(n)$ gauge theory with no matter multiplets 
is conjectured to be the moduli space of $n$ BPS monopoles 
of $SU(2)$ gauge group. This proposal was first made in \cite{sw3} 
for the case $n=2$ and later generalised in \cite{chahan}. 
The case $n=2$ has subsequently been proven by explicit field 
theory calculations \cite{dkmtv}. For the case of general $n$ the 
equivalence has been used to predict the leading order exponential 
corrections to the $n$-monopole $SU(2)$ moduli space \cite{cd}. The 
connection between monopoles and three-dimensional gauge theories 
appears most naturally in a D-brane setting \cite{hw}. Moreover, 
this technology provides a further generalisation; the Coulomb branches 
of $N=4$ three dimensional gauge theories with product gauge groups and 
certain bi-fundamental matter couplings are conjectured to be the moduli 
spaces of BPS monopoles of $SU(r+1)$ gauge groups for any $r\geq 1$. 

In the following we propose to extend this correspondence yet further. 
We construct product gauge groups with matter content determined by 
ADE Dynkin diagrams. We show that one-loop corrections to the metrics on 
the Coulomb branches reproduce the asymptotic metrics on the moduli spaces of 
BPS monopoles of any simply-laced Lie group. We conjecture that this 
correspondence is exact.

Note added: While writing this paper, reference \cite{sosp} 
appeared in which the authors construct the moduli space of 
$SO$ and $Sp$ monopoles from branes with an orientifold plane. 
It would be interesting to examine the corresponding three-dimensional 
gauge theories in these cases.

\subsubsection*{\em Three-Dimensional Gauge Theories}

Three-dimensional $N=4$ supersymmetric theories are the dimensional reduction 
of the minimal supersymmetric models in six dimensions. Representations 
of the supersymmetry algebra come as either vector or hyper multiplets. 
The former contains a three-dimensional gauge field, $A_\mu$, 
three real scalars which we write in vector form, $\vec{\phi}$, 
and four Majorana fermions. All fields in the vector multiplet transform
 in the adjoint representation 
of the gauge group. Matter fields are contained in hypermultiplets which 
consist of  four Majorana fermions, now paired with four real scalars. 
We will only consider hypermultiplets transforming in the (anti)-fundamental 
representation of the gauge group. The $N=4$ action has a 
$SO(3)_N\times SO(3)_R$ global R-symmetry under which all fields transform. 
In particular, $\vec{\phi}$ transforms in the ${\bf 3}$ of $SO(3)_N$ and is 
invariant under $SO(3)_R$. 

We will denote the three-dimensional gauge group as ${\cal G}$. 
It is a product group constructed from the Dynkin diagram of an auxiliary 
simply laced group, $G$, of rank $r$. Let $\ubeta^A$, $A=1,..,r$ be the 
simple roots of $G$ normalised as $\ubeta^A\cdot\ubeta^A=1$. To each 
assign a positive integer, $n_A$.  To the node of the Dynkin diagram 
corresponding to the simple root $\ubeta^A$, we associate the 
three-dimensional gauge group 
$U(n_A)$ with coupling constant $e_A$. Thus ${\cal G}$ is of 
rank $R=\sum_An_A$. We also include matter content, ${\cal C}$. 
For each link in the Dynkin diagram 
connecting the $\ubeta^A$ node with the $\ubeta^B$ node, we add a 
hypermulitplet transforming in the bi-fundamental representation 
$({\bf n}_A,\bar{\bf n}_B)$. Thus the field content is,     
\begin{equation}\label{gg}
{\cal G}=\bigotimes_{A=1}^rU(n_A)\hspace{1cm};\hspace{1cm}
{\cal C}=\bigoplus_{A\neq B}a_{AB}({\bf n}_A,\bar{\bf n}_B)\, ,
\end{equation}
where $a_{AB}=1$ whenever $\ubeta^A\cdot\ubeta^B\neq 0$ and is zero 
otherwise. For $A_n$ Dynkin diagrams, this construction coincides 
with the brane picture of \cite{hw}. The cases of $D_n$ and exceptional 
Dynkin diagrams are new.

The construction above is similar to Kronheimer's hyper\Ka quotient 
construction of ALE spaces \cite{kron}. These were studied in the context 
of three-dimensional $N=4$ theories in \cite{is} (see also \cite{berk} 
for related quiver constructions) where the Coulomb branch 
was proposed to be the moduli space of instantons of gauge group $G$. The 
constructions differ in the use of the extended Dynkin diagram in 
\cite{kron} where the integers $n_A$ are also set equal to the 
Dynkin indices. 
  
On the Coulomb branch, the hypermultiplet scalars 
have zero vacuum expectation value (VEV) while the VEVs of the vector 
multiplet scalars are constrained to live in the $R$-dimensional 
Cartan subalgebra (CSA), ${\bf H}$, of ${\cal G}$. 
\begin{equation}\label{vev}
\langle\vec{\phi}\rangle =\vec{\bf v}\cdot{\bf H}\, .
\end{equation}
We assume the VEVs break ${\cal G}$ to the 
maximal torus. The adjoint Higgs mechanism gives a mass to all fields 
except vector multiplet fields in ${\bf H}$. We denote the 
massless gauge fields as ${\bf A}_\mu={\rm Tr}(A_\mu{\bf H})$ 
with similar notation for the other fields. The number 
of massless bosonic fields is thus $3R$ real scalars and 
$R$ abelian gauge fields. 

We consider the Wilsonian low-energy effective action for the 
massless degrees of freedom. Introduce $R$ orthonormal vectors, 
${\bf e}_i$, $i=1,..,R$, spanning the root space of ${\cal G}$. 
In order to later compare with the monopole  moduli space it will prove 
useful to introduce $r$ new numbers 
$t_B=\sum_{\mbox{\tiny $A=1$}}^{\mbox{\tiny $B-1$}}n_A$ 
where we set $t_1=1$. The vectors ${\bf e}_i$, 
$t_A\leq i<t_{A+1}$ are associated with the CSA of the $U(n_A)$ factor of 
the gauge group. We also define the coupling constant $e_i=e_A$ 
for $t_A\leq i<t_{A+1}$.

At tree level, the bosonic sector of the low-energy effective action 
is a free abelian theory,
\begin{eqnarray}\label{claslow}
S_B&=&\sum_{i,j=1}^R\,\int{\rm d}^3x
\,{\cal M}_{ij}\left(-\ft14 ({\bf F}\cdot{\bf e}_i)({\bf F}\cdot{\bf e}_j
)+\ft12({\partial\vec{\bphi}}\cdot{\bf e}_i)\cdot(\partial{\vec{\bphi}}
\cdot{\bf e}_j)\right)\, ,
\end{eqnarray}  
where 
\begin{equation}\label{Mclas}
{\cal M}_{ij}=\frac{2\pi}{e_i^2}\delta_{ij}\, .
\end{equation}
We have supressed all space-time indices, and $F\equiv F_{\mu\nu}$ 
is the abelian field strength.

At weak coupling the low-energy effective action receives contributions 
from perturbation theory and instantons. We employ the background field 
method to calculate the former at one-loop. Details of the calculation for 
$SU(2)$ gauge group can be 
found in appendix B of \cite{dkmtv} and the appendix of \cite{dtv}. The 
extension to arbitrary gauge group is simple. Firstly consider the 
effects of integrating out the vector multiplets. Denote 
the roots of $U(n_A)$ as $\blambda^A_m$, $m=1,..,N_A=n_A(n_A-1)$. 
Explicitly, $\{\blambda^A_m\,;\,m=1,..,N_A\}\equiv\{{\bf e}_i-{\bf e}_j\,;
\,i,j=t_A,..,t_A-1\}$. For each factor $U(n_A)$ of the gauge group, 
${\cal G}$, the low-energy effective action includes the term
\begin{equation}\label{vecpert}
\frac{1}{16\pi}\sum_{m=1}^{N_A}\,\left(\frac{({\bf F}\cdot
\blambda_m^A)^2-2|\partial{\vec{\bphi}}\cdot\blambda^A_m)|^2}
{|{\vec{\bf v}}^A\cdot\blambda^A_m|}\right)\, .
\end{equation}

For the hypermultiplets, denote the weights of the 
fundamental representation of $U(n_A)$ as ${\bf w}^A_p$, $p=1,..,n_A$. 
The weights of the anti-fundamental are $-{\bf w}^A_p$. We have 
$\{{\bf w}^A_p\,;\,p=1,..,n_A\}\equiv\{{\bf e}_i\,;\,i=t_A,..,t_A-1\}$. 
Integrating out 
high momentum modes of the hypermultiplet transforming in the 
$({\bf n}_A,\bar{\bf n}_B)$ representation yields
\begin{equation}\label{hyppert}
-\frac{1}{16\pi}\sum_{p=1}^{n_A}\sum_{q=1}^{n_B}\, \left(\frac{(
{\bf F}\cdot{\bf w}^A_p-{\bf F}\cdot{\bf w}^B_q)^2-
2|\partial\vec{\bphi}\cdot{\bf w}^A_p-\partial\vec{\bphi}
\cdot{\bf w}^B_q|^2}
{|\vec{\bf v}\cdot{\bf w}^A_p-\vec{\bf v}\cdot{\bf w}^B_q|}\right)\, .
\end{equation}
Equations \eqn{vecpert} and \eqn{hyppert} can both be interpreted as 
coupling constant renormalisations. There are further one-loop corrections 
to the low-energy effective action which are not of this form \cite{seib}. 
These will be dictated by supersymmetry.

Combining equations, \eqn{claslow}, \eqn{vecpert} and \eqn{hyppert}, 
the low-energy effective action can be written in the form \eqn{claslow}, 
with 
\begin{eqnarray}\label{Mpert}
{\cal M}_{ii}&=&\frac{2\pi}{e_i^2}-\frac{1}{2\pi}
\sum_{k\neq i}\frac{\ualpha_i\cdot\ualpha_k}{|\vec{\bf v}\cdot 
({\bf e}_i-{\bf e}_k)|} \nonumber \\
{\cal M}_{ij}&=&\frac{1}{2\pi}\frac{\ualpha_i\cdot\ualpha_j}{|\vec{\bf v}
\cdot ({\bf e}_i-{\bf e}_j)|}
\end{eqnarray}
where we have defined $\ualpha_i=\ubeta_A$ for $t_A\leq i<t_{A+1}$ 
(recall $\ubeta^A$ are the simple roots of the auxillary simply-laced group, 
$G$). 

The $R$ abelian gauge fields, ${\bf A}_\mu$, are dual to scalars 
$\bsigma$ which serve as Lagrange multipliers for the Bianchi identity. 
We add to the action the surface term 
\begin{equation}\label{surface}
S_S=\frac{i}{4\pi}\sum_{i=1}^R\,\int{\rm d}^3x\,\epsilon^{\mu\nu\rho}
(\partial_\mu{\bf F}_{\nu\rho}\cdot{\bf e}_i)(\bsigma\cdot{\bf e}_i)
\end{equation}
where we have restored the space-time indices. With this normalisation, 
the scalars $\bsigma\cdot{\bf e}_i$ have period $2\pi$ in the 
background of a single instanton. Performing the Gaussian 
integration over ${\bf F}_{\mu\nu}$  we promote ${\bf \sigma}$ to a full 
dynamical field. The low-energy effective action is now a $\sigma$-model 
with coordinates $\vec{\bphi}\cdot{\bf e}_i$ and $\bsigma\cdot{\bf e}_i$ 
on the 4$R$-dimensional target space,
\begin{equation}\label{sigpert}
S_{\rm 1-loop}=-\frac{1}{2}\sum_{i,j=1}^R\,\int\,{\rm d}^3x\,{\cal M}_{ij}
(\partial{\vec{\bphi}}\cdot{\bf e}_i)\cdot({\partial\vec{\bphi}}
\cdot{\bf e}_j)+\frac{1}{4\pi^2}({\cal M}^{-1})_{ij}
(\partial\bsigma\cdot{\bf e}_i)(\partial\bsigma\cdot{\bf e}_j)\, .
\end{equation}
This action inherits the $SO(3)_N$ symmetry of the microsocpic action. 
It also posseses $R$ global $U(1)$ symmetries, 
\begin{equation}\label{sigshift}
{\bsigma}\rightarrow\bsigma+{\bf c}
\end{equation}
for any constant $R$-vector, ${\bf c}$. Although broken by instanton 
effects, these symmetries exist to all orders in perturbation theory as 
can be seen by integrating \eqn{surface} by parts in a topologically 
trivial background.  

The global symmetries of the action translate to isometries 
of the metric on the target space. Moreover, $N=4$ supersymmetry requires 
that this metric is hyper\Ka \cite{AG}. $4R$-dimensional hyper\Ka 
metrics with $R$ triholomorphic $U(1)$ isometries have a simple form 
\cite{Pedpoon}. The hyper\Ka condition requires augmenting the 
bosonic action \eqn{sigpert} with extra terms generated at one-loop, 
corresponding to the replacement, 
\begin{equation}\label{hkadd}
\partial\bsigma\cdot{\bf e}_i\rightarrow\partial\bsigma\cdot{\bf e}_i
+\sum_{k=1}^R\vec{\cal W}_{ik}\cdot({\partial\vec{\bphi}}\cdot{\bf e}_k)\, ,
\end{equation}
where,
\begin{equation}
\vec{\nabla}\times\vec{\cal W}_{ij}=-2\pi\vec{\nabla}{\cal M}_{ij}\, .
\end{equation}
The derivative, $\vec{\nabla}$, is taken with respect to 
$\vec{\bf v}\cdot{\bf e}_i$. Equations \eqn{sigpert} and \eqn{hkadd} 
define the bosonic one-loop low-energy effective action. It remains to show 
that this is equivalent to a sigma model on a monopole moduli space.

\subsubsection*{\em Monopole Moduli Spaces}

We consider moduli spaces of BPS monopoles of the simply-laced gauge groups 
$G$. An adjoint Higgs field with vanishing potential 
is assumed to break $G$ to the maximal torus, and in doing so defines 
$r$ simple roots, $\ubeta_A$, $A=1,..,r$.  The theory contains 
monopole configurations with magnetic charge defined by an $r$-dimensional 
vector, ${\ug}$. Topological considerations force ${\ug}$, to lie 
in the root lattice (with suitably normalised roots) of $G$ and we expand 
${\ug} =\sum_An_A\ubeta_A$. 

The moduli space of such a monopole configuration has 
dimension $4R=4\sum_A n_A$ \cite{wein}. This result has the interpretation 
that there exist $r$ types of ``fundamental'' monopoles corresponding 
to magnetic charges ${\ug} =\ubeta_A$ for $A=1,..,r$. The only moduli of 
a fundamental monople configuration are the position, ${\vec x}$, 
and the phase, $\xi^A$, generated by global 
gauge transformations within the maximal torus. A general monopole 
configuration can be thought of as consisting of $R$ individual 
fundamental monopoles, $n_A$ of each type, at least in the asymptotic 
region of the moduli space. Let the $i^{\rm th}$ monopole be 
associated with the simple root, $\ualpha_i=\ubeta_A$ for some $A$. 
For well seperated monopoles, we may ascribe positions, ${\vec x}_{i}$, 
and phases $\xi_{i}$ to each monopole. We also define 
${\vec r}_{ij}={\vec x}_i-{\vec x}_j$ and $r_{ij}=|{\vec r}_{ij}|$. 

The full metric on the monopole moduli space is not known. 
In the asymptotic region the monopoles are well-seperated and 
the metric may be calculated using the techniques of \cite{man2,gm}. The 
calculation for arbitrary gauge group was performed in \cite{lwy} and 
the metric has the form\footnote{In the notation of \cite{lwy} we set 
$g=2$ where $g$ is the coupling constant for monopole interactions. 
This parameter has no counterpart in the three-dimensional theory},
\begin{equation}\label{monmetric}
{\rm d}s^2 = M_{ij}{\rm d}{\vec x}^{i}\cdot
{\rm d}{\vec x}^{j}+\frac{1}{4\pi^2}(M^{-1})_{ij}({\rm d}\xi^{i}+{\vec W}_{ik}
\cdot{\rm d}{\vec x}^{k})({\rm d}\xi^{j}+{\vec W}_{jl}\cdot{\rm d}{\vec x}^{l})
\end{equation}
where,  
\begin{eqnarray}
M_{ii}&=&m_i-\sum_{k\neq i}
\frac{\ualpha_i\cdot\ualpha_k}{2\pi r_{ik}}\nonumber \\ 
M_{ij}&=&\frac{\ualpha_i\cdot\ualpha_j}{2\pi r_{ij}} 
\end{eqnarray}

$m_i$ is the mass of the $i^{\rm th}$ fundamental monopole associated 
with the simple root $\ualpha_i$ and ${\vec W}_{ij}$ is defined by 
$\vec{\nabla}\times\vec{W}_{ij}=-2\pi\vec{\nabla}M_{ij}$. 

Comparing \eqn{monmetric} to the sigma model metric defined by 
\eqn{sigpert} and \eqn{hkadd} we find the metrics do indeed 
coincide as advertised with
\begin{eqnarray}\label{matching}
m_i&=&2\pi/e^2_i \nonumber \\
\vec{x}_i&=&{\vec{\bphi}}\cdot{\bf e}_i \nonumber \\
\xi_i&=&\bsigma\cdot{\bf e}_i 
\end{eqnarray}
The metric \eqn{monmetric} is singular at $r_{ij}=0$ for $\ualpha_i\cdot
\ualpha_j>0$. These singularities are resolved by exponential corrections 
to the metric corresponding to instanton corrections in the 
three-dimensional gauge theory. Assuming the correspondence between monopole 
moduli spaces and three-dimensional gauge theories is exact, the leading 
order exponential corrections could be calculated along the lines of 
\cite{cd}.

\vspace{.5in}

\centerline{*******************}

I am grateful to N. Dorey and S. Vandoren for valuable discussions and 
to the University of Washington for hospitality. I am supported by PPARC.

\end{document}